\documentclass[runningheads]{cl2emult}

\usepackage{makeidx}  
\usepackage{graphicx} 
\usepackage{subeqnar} 
\usepackage{multicol} 
\usepackage{cropmark} 
\usepackage{eso}      
\makeindex            

\def\etal{{\it et al.} }

\def\apj{ApJ}

\def\aa{{\it A\&A}}

\def\inprep{in preparation}

\def\cm2{$cm^{-2}$}

\begin{document}
\title*{How to make CMB maps from huge timelines with small computers}
\toctitle{How to make CMB maps from huge timelines with small computers}
%
%
\titlerunning{How to make CMB maps...}

\author{Xavier Dupac
\and Martin Giard}

\authorrunning{Dupac \& Giard}
\institute{Centre d'\'Etude Spatiale des Rayonnements \\
9, avenue du Colonel Roche \\
BP 4346 \\
31028 Toulouse cedex 4 \\
France}

\maketitle              

\begin{abstract}

We present in this article two different ways to make CMB maps in practice,
from large timelines.
One is to make a simple destriping, fitting the data and using the scan
intercepts to remove the low frequency noise (stripes).
The second, optimal, is to resolve linearly the map-making problem, which in
case of big timelines must be simplified and changed from matrices to vectors
for the calculations.
Assuming few conditions on the noise, it is possible to make fast map-making tools.

\end{abstract}

\section{Introduction}

The last ten and the future ten years provide a large number of CMB (Cosmic
Microwave Background) experiments.
Their main goal is to estimate the
cosmological parameters through sky temperature maps.
The map-making is therefore a key question of these experiments, moreover with
the advent of large time-ordered data (TOD) being currently analysed or simulated.
The "brute force" direct inversion of the linear map-making problem is beyond
the reach of computing facilities of today.
Thus the data analysers need to find faster methods, which have to be as
optimal in theory and as efficient in practice as possible.

\section{Destriping for Planck HFI data}

This is used to clean Planck High Frequency Instrument (Bersanelli \etal 1996) simulated data, to be published in
Giard \etal (2000).
It uses a simple algorithm to destripe the data and make the map.
The scanning strategy assumed to make the timeline is 1 round per minute of
the beam on the sky, so that the timeline is a succession of circles on the
sky.
The cleaning algorithm follow the processing described below:
First each circle is adjusted so that the measured signal fits a cosecant law
(galactic dust), plus dipole emission:

\begin{equation}
M_{ik} = r_k (Dipole_{ik} + \frac{g_k}{sin(bII_i)} + C^0_k)
\end{equation}

where k is the index of each circle of the data, i the data index along the
circle, and $r_k$ the detector response.
The free parameters are $g_k$ (galactic emission for the circle k) and $C^0_k$
(constant to be subtracted).
Then we destripe the data using an algorithm derived from Delabrouille (1998)
which uses the scan intercepts.
We adjust the constants to subtract to each circle k by minimising the spread
between the measurements contributing to the same sky pixel.

\section{The optimal method in practice}

\subsection{The map-making problem}

If we assume the instrument to be linear, the time-ordered data, represented
by a vector y, is linearly
dependent on the real sky, represented
by a vector x.
We have thus:

\begin{equation}
y = A x + n
\end{equation}

where A is the point-spread (convolution) matrix and n the random noise vector
in the timeline.
For CMB experiments, x would represent the pixelised sky map of the CMB temperature.
The map-making problem is thus written as this:

\begin{equation}
x_{0} = W y
\label{deconv_eq}
\end{equation}

where $x_{0}$ is the vector of the reconstructed sky map, and W the inversion
matrix.
The inversion matrix W depends on the method used. The simplest
is the pixel averaging, but
the optimal methods for estimating the map are the
COBE method (with no prior for x) or the Wiener filter (with gaussian prior
for x).
See Tegmark (1997) for a set of linear and non-linear map-making methods.
As an example, the COBE method is written as this:

\begin{equation}
W = [A^{t} N^{-1} A]^{-1} A^{t} N^{-1}
\label{cobe}
\end{equation}

with N the noise covariance matrix $N = <n n^{t}>$.

\subsection{The large timeline problem}

The CMB experiments now in analysis (Boomerang, MAXIMA...) or to come
(Archeops, MAP, Planck...) provide several megabytes of time-ordered
data.
The noise covariance matrix, for instance, would be several terabytes of data, which is
impossible to handle or even to write.
How to solve optimally the map-making problem without writing any matrix
anywhere ?

Assuming no beam and the stationarity of the noise, it is possible to make the
map iteratively handling only vectors.
Using the COBE method (eq. \ref{cobe}), the deconvolution equation (eq. \ref{deconv_eq}) can be written as this:

\begin{equation}
[A^{t} N^{-1} A] x_{0} = A^{t} N^{-1} y
\label{eqltm}
\end{equation}

If the noise is stationary, the matrix multiplication $N^{-1}$ y is equivalent to a convolution by a kernel, i.e. a
multiplication in the Fourier space.
If the beam is not taken into account, A is making a timeline from a map, and
$A^{t}$ is the pixel averaging of a timeline into a map.
Eq. \ref{eqltm} leads to iterative methods such as:

\begin{equation}
x_{n+1} = x_n + \eta A^{t} N^{-1} y - \eta A^{t} N^{-1} A x_n
\end{equation}

where $\eta$ is a free parameter of the iterative method.
It is also possible to make converge the noise map instead of directly the sky
map, by changing the variable $x_0$ into $nx_0 = x_0 - [A^t A]^{-1} A^t y$.

\begin{figure}
\centering
\includegraphics[width=.4\textwidth]{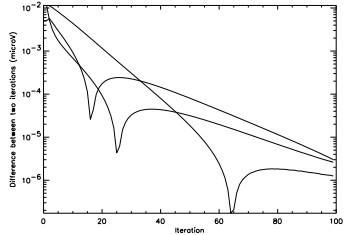}
\caption[]{Convergence for three pixels of Archeops technological flight data}
\label{precision}
\end{figure}

Fig. \ref{precision} shows the convergence of such methods, here applied to
data from Archeops technological flight, realised at Trapani
in sept. 99 (Beno\^\i t \etal 2000).

\clearpage
\addcontentsline{toc}{section}{Index}
\flushbottom
\printindex

\end{document}